\documentclass[aps,prb,twocolumn,floatfix,showpacs]{revtex4}
\usepackage{graphicx}
\usepackage{bbm}
\usepackage{amsmath,amssymb}
\newcommand{\be}{\begin{equation}}
\newcommand{\beq}{\begin{equation}}
\newcommand{\ee}{\end{equation}}
\newcommand{\bea}{\begin{eqnarray}}
\newcommand{\eea}{\end{eqnarray}}
\newcommand{\ba}{\begin{array}}
\newcommand{\ea}{\end{array}}

\begin{document}
\title{Ultrafast sequential charge transfer in a double quantum dot}
\author{A. Putaja}
\affiliation{Nanoscience Center, Department of Physics, University of
  Jyv\"askyl\"a, FI-40014 Jyv\"askyl\"a, Finland}
\author{E. R{\"a}s{\"a}nen}
\email[Electronic address:\;]{erasanen@jyu.fi}
\affiliation{Nanoscience Center, Department of Physics, University of
  Jyv\"askyl\"a, FI-40014 Jyv\"askyl\"a, Finland}

\date{\today}

\begin{abstract}
We use optimal control theory to construct external electric fields which 
coherently transfer the electronic charge in a double quantum-dot system.
Without truncation of the eigenstates we operate on desired superpositions 
of the states in order to prepare the system to a localized state and to
coherently transfer the charge from one well to another.
Within a fixed time interval, the optimal processes are shown to occur 
through several excited states. The obtained yields are generally 
between $99\,\%$ and $99.99\,\%$ depending on the field constraints, and
they are not dramatically affected by strict frequency filters 
which make the fields (e.g., laser pulses) closer to experimental realism.
Finally we demonstrate that our scheme provides simple access to hundreds of 
sequential processes in charge localization while preserving the high fidelity.
\end{abstract}


\maketitle

\section{Introduction}

During the past few years, coherent control of charge in
double quantum dots (DQDs) has been a subject of 
active experimental~\cite{petta, petta2, gorman, kataoka} 
and theoretical~\cite{forre1,kosionis,2D_DQD,salen,nepstad,fountoulakis}
research.
Here one of the long-term aims is the design of a
solid-state quantum computing scheme.~\cite{loss}
It is still to be seen whether the optimal control mechanism
DQDs turns out to operate through magnetic fields,~\cite{popsueva,harju}
gate voltages,~\cite{kataoka} or optimized laser pulses.~\cite{2D_DQD,nepstad}

Dynamical control of charge in DQDs has been a popular
application for few-level 
schemes~\cite{openov,brandes,grigorenko,paspalakis,voutsinas,selsto,kosionis,fountoulakis}
(modeling DQDs as two-, three-, or four-level systems), which have demonstrated ultrafast 
high-fidelity processes. However, a physical
DQD has, in principle, infinitely many levels, 
and in fast processes a considerable number of
states might have practical relevance. 
For example, a two-level approximation is {\em exact} only in the limit of
using an infinitely long resonant continuous wave with an infinitely 
small amplitude. A linear field (bias) is an appealing and simple
alternative to control charge in DQDs,~\cite{forre1} 
but it is not applicable to fast processes. 

With quantum optimal control theory~\cite{oct,janreview} (OCT) it is possible
to find optimized external fields driving the system  -- having
an arbitrary number of states -- from the initial state to the 
desired target state without any
approximations, apart from a possible model potential 
to describe the physical apparatus. OCT has been
used to analyze the general controllability criteria 
of two-dimensional single-electron DQDs and to optimize interdot 
charge transfer.~\cite{2D_DQD} Optimal control of two-electron
DQDs has been obtained in an extensive work of
Nepstad {\em at al.}~\cite{nepstad} addressing various 
control schemes~\cite{degani} and hyperfine interactions.~\cite{hyperfine}

In this work we apply OCT to construct external electric fields that lead to
fast sequential charge transfer processes in single-electron DQDs. To obtain 
high fidelity we operate on the superpositions of the lowest
states corresponding to the charge localization in left or 
right well. We show that hundreds of sequential charge
transfer processes can be achieved without a significant loss
of the yield. To make the experimental production of the obtained fields more 
realistic, we cut off the high-frequency components already during 
the optimization procedure. The use of such filters does not dramatically
affect the fidelity.

\section{Model}\label{model}

\begin{figure}
\includegraphics[width=0.70\columnwidth]{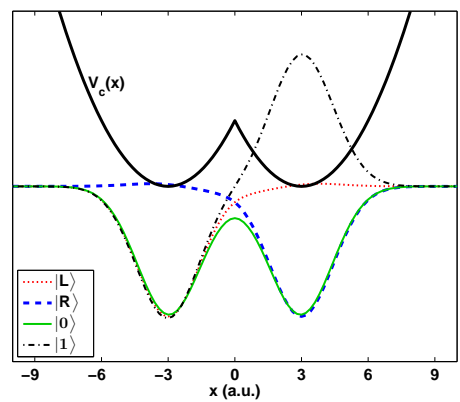}
\caption{(Color online) Model potential for the quantum dot (black solid line), 
the ground state (green line), the first excited state (black dash-dotted line), and their
superpositions corresponding to left (red dotted line) and right (blue dashed line) states.}
\label{fig1}
\end{figure}

We use a one-dimensional (1D) model describing a single-electron 
semiconductor DQD. The external potential has a form
\begin{equation}\label{vc}
 V_{c}(x)=\frac{\omega_{0}^{2}}{2}\,\textrm{min}\Bigl\{\Bigl(x-\frac{d}{2}\Bigr)^{2},\Bigl(x+\frac{d}{2}\Bigr)^{2}\Bigr\}
\end{equation}
in effective atomic units (a.u.), see below.
Here $d=6$ is the interdot distance and $\omega_{0}=0.5$ is the confinement strength. The potential is visualized in Fig.~\ref{fig1}.
We consider typical 
GaAs material parameters within the effective-mass approximation, i.e.,
$m^*=0.067$ and $\epsilon=12.7$. Now, the energies, lengths, and times scale as 
$E_{h}^\ast=(m^\ast/m_0)/(\varepsilon/\varepsilon_0)^2E_h\approx 11\,\mathrm{meV}$,
$a_0^\ast=(\varepsilon/\varepsilon_0)/(m^\ast/m_0) a_0\approx 10\,\mathrm{nm}$, and
$t_0^\ast=\hbar/E_h^\ast\approx 60\,\mathrm{fs}$, respectively. We emphasize
that below the abbreviation a.u. refers to these effective atomic units.

It should be noted that a harmonic potential in Eq.~(\ref{vc}) is,
in its two-dimensional (2D) form, a general model for realistic 
semiconductor quantum-dot structures.~\cite{qd_review} Since the first 
Coulomb-blockade experiments it has been shown that the harmonic model
is essentially valid up to dozens of electrons confined in the dot, 
and thus up to a large number of levels. The validity is clear, e.g.,
in recent works combining experiments and theory in the spin-blockade 
regime.~\cite{spindroplet,rogge} The precise energy-level spectrum
in a given device can be explicitly obtained through single-electron 
transport experiments, and this information can be utilized to 
reconstruct the particular form of the extenal potential. 
For example, in Ref.~\onlinecite{impurity} it was explicitly shown that 
measured energy-level spectrum can be well reproduced by a harmonic model 
potential upon slight refinements. Hence, when necessary, Eq.~(\ref{vc})
can be tuned to match a particular device. Regarding the results below, 
the 1D model does not yield a qualitative difference from a more 
realistic 2D potential, but it significantly speeds up the calculations.

Electronic states localized to left and right dots can be expressed 
as superpositions of
the two lowest (gerade and ungerade) 
states $\left.|0\right>$ and $\left.|1\right>$ as follows:
\begin{eqnarray}
\left.|L\right> & = & \frac{1}{\sqrt{2}}(\left.|0\right>+\left.|1\right>)\\
\label{states}
\left.|R\right> & = & \frac{1}{\sqrt{2}}(\left.|0\right>-\left.|1\right>)      
\end{eqnarray}
If the system is prepared in either of the superpositions, the occupation probabilities 
of $\left.|L\right>$ and $\left.|R\right>$ oscillate 
with the resonance frequency $\omega_{01}=E_{1}-E_{0}\approx 0.0135$ (see Ref.~\onlinecite{sakurai}).
For instance, if the system is first prepared at $\left.|L\right>$, 
it reaches the state $\left.|R\right>$ at $t=T/2=\pi/\omega_{01}\approx 232.87$.
As discussed in detail below, we aim at {\em controlling} this 
charge-transfer procedure in an arbitrary way.

\section{Method}

In OCT the objective is to find an external time-dependent field 
$\boldsymbol{\epsilon}(t)$ that drives the system into the predefined state through
the solution of the Schr\"odinger equation,
\begin{equation}\label{schr}
i\frac{\partial}{\partial t}\Psi(\boldsymbol{r},t)=\hat{H}[\boldsymbol{\epsilon}_{k}(t)]\Psi(\boldsymbol{r},t).
\end{equation}
Here $\boldsymbol{\epsilon}(t)$ is an electric field (e.g., laser pulse) 
dealt with the dipole approximation, so that the Hamiltonian has the form,
\begin{equation}
\hat{H} = \hat{T}+\hat{V}_c-\hat{\mu}\boldsymbol{\epsilon}(t),
\end{equation}
where the (static) external potential is that of Eq.~(\ref{vc})
where $\hat{\mu}=-{\mathbf r}$ is the dipole operator.

Starting with an initial guess for the electric field $\boldsymbol{\epsilon}(t)$,
we maximize the expectation value of the target operator $\hat{O}$:
\begin{equation}
\label{j1}
 J_{1}[\psi]=\left<\Psi(\boldsymbol{r},T)|\hat{O}|\Psi(\boldsymbol{r},T)\right>,
\end{equation}
where $\hat{O}=\left.|\Phi_{\rm F}\right>\left<\Phi_{\rm F}|\right.$ is now a 
projection operator, since we aim at maximizing the occupation of the target state $\Phi_{\rm F}$
at the end of the field at time $T$:
\begin{equation}
 J_{1}=\left|\big<\Psi(\boldsymbol{r},T)|\Phi_{\rm F}\big>\right|^{2}.
\end{equation}
In the following, this quantity is referred to the {\em yield}.

As a constraint, avoiding fields with very high energy, the fluence 
(time-integrated intensity) of the field is limited by a second functional,
\begin{equation}
 J_{2}[\epsilon]=-\alpha\left[\int_{0}^{T}dt\,\epsilon^{2}(t)-E_0\right],
\end{equation}
where $E_0$ is the fixed fluence [see Eq.~(\ref{controleqs}) below]
and $\alpha$ is a time-independent Lagrange multiplier.~\cite{janreview}

Finally, the satisfaction of the time-dependent Schr\"odinger 
equation [Eq.~(\ref{schr})] introduces 
yet another functional,
\begin{equation}
  J_{3}[\epsilon,\Psi,\chi] = -2\,\textrm{Im}\int_{0}^{T}\big<\chi(t)|i\partial_{t}-\hat{H}(t)|\Psi(t)\big>,
\end{equation}
where $\chi(t)$ is a time-dependent Lagrange multiplier.

Variation of $J=J_{1}+ J_{2}+J_{3}$ with respect to $\Psi$, $\chi$, $\epsilon$, and $\alpha$ lead 
to the {\em control equations} 
\begin{eqnarray}
	i\partial_{t} \Psi(t) & = & \hat{H}(t)\Psi(t), \quad \Psi(0)=\Phi_{I},\\
	i\partial_{t} \chi(t) & = & \hat{H}(t)\chi(t), \quad \chi(T)=\hat{O}\Psi(T),\\
	\epsilon(t) & = & -\frac{1}{\alpha}\textrm{Im}\big<\chi(t)|\mu|\Psi(t)\big>,\\
        \int_{0}^{T}dt\,\epsilon^{2}(t) & = & E_0.\label{controleqs}
\end{eqnarray}
which can be solved iteratively.~\cite{zhu, janreview} 
We apply a numerically
efficient forward-backward propagation scheme introduced by Werschnik and 
Gross.~\cite{Werschnik2}
When solving the control equations, the Lagrange multiplier
$\alpha$ is calculated through the fixed fluence $E_0$ as explained in detail
in Ref.~\onlinecite{janreview}. The field is constrained by an envelope function
of a form
\begin{equation}
f(t) = \frac{1}{2}\left\{\mbox{Erf}\left[\frac{a}{T}\left(t-\frac{T}{b}\right)\right]+ \mbox{Erf}\left[-\frac{a}{T}\left(t-T+\frac{T}{b}\right)\right]\right\}
\end{equation}
with $a=100$ and $b=20$. This corresponds to a step function ascending (descending) rapidly
at $t\sim b/4$ ($t\sim T-b/4$).
The scheme also allows straightforward inclusion of spectral 
constraints discussed in the following section. 
In the numerical calculations we have used the {\tt octopus} code~\cite{octopus}
which solves the control equations in real time on a real-space grid.

To approximate the time-propagator we applied the time-reversal symmetry,
i.e., propagating $\Psi(t)$ forward by $\Delta t/2$ should correspond to
propagating $\Psi(t+\Delta t)$ backward by $\Delta t/2$. This condition
leads to an approximation for the propagator,~\cite{evolution}
which can be further improved by extrapolating the time-dependent potentials.
In {\tt octopus}~\cite{octopus} the used method is called 
Approximated Enforced Time-Reversal Symmetry (AETRS).

\section{Results}

First, the system is {\em prepared} from the ground state $\left.|0\right>$ to the 
desired superposition. Hence, we simply define the target wave function in Eq.~(\ref{j1})
as $\left.|\Phi_{\rm F}\right>=\left.|L\right>$. 
We set the field length to $T=100$ ($\sim 6$ ps) and the initial frequency 
to $0.5$ corresponding to the oscillator frequency $\omega_0$ of the DQD. 
Unless stated 
otherwise, the fluence is fixed to $E_0=0.3$, so that the average intensities are of the
order of $10^3$ W/cm$^2$ (note the units given in Sec.~\ref{model}). 

The optimized field in Fig.~\ref{fig1}(a) 
\begin{figure}
\includegraphics[width=0.80\columnwidth]{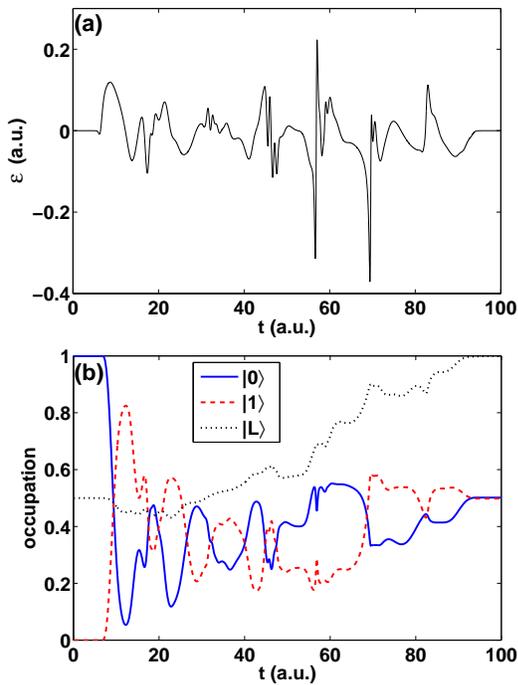}
\caption{(Color online) (a) Optimized field to prepare the system to the
superposition $\left.|L\right>$ (see text) from the ground state.
(b) Occupations of the ground state $\left.|0\right>$, first excited
state $\left.|1\right>$ and their superposition $\left.|L\right>$.
}
\label{fig2}
\end{figure}
looks rather complicated with
distinct high-frequency components, whose role and possible removal 
is discussed in detail below. The occupations of the states, i.e.,
their overlaps with the time-propagated wave function, are plotted
in Fig.~\ref{fig1}(b). The ground state (initially occupied) and
the first excited state (initially empty) get half populated, so that
their superposition $\left.|L\right>$ becomes fully populated and
the electron is localized in the left well. The obtained yield
is as high as $0.99985$.

\begin{figure}
\includegraphics[width=0.85\columnwidth]{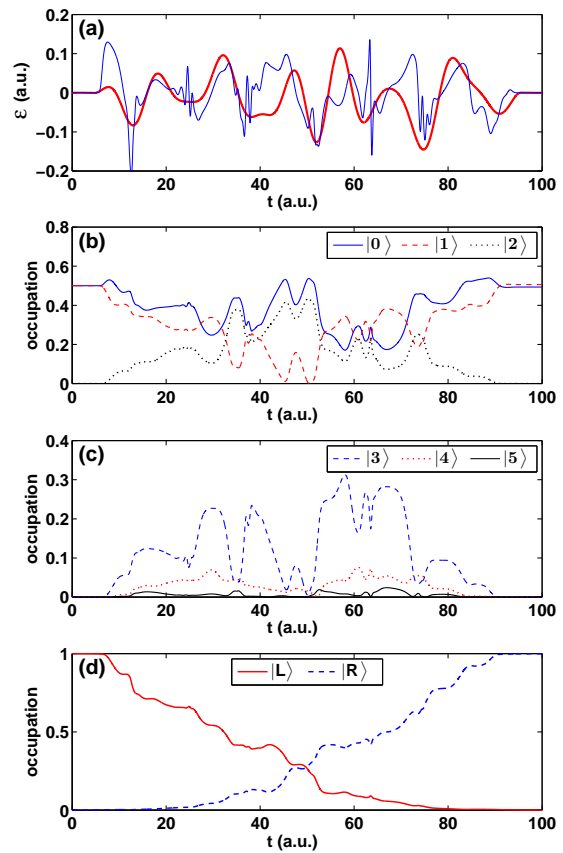}
\caption{(Color online) (a) Optimized fields without (thin blue line) and with spectral constraints (thick red line) 
for transition $\left.|L\right>\rightarrow\left.|R\right>$. (b-c) Occupations of the five 
lowest eigenstates during the process. (d) Occupations of the initial and
target superposition states.}
\label{fig3}
\end{figure}

\begin{figure}
\includegraphics[width=0.70\columnwidth]{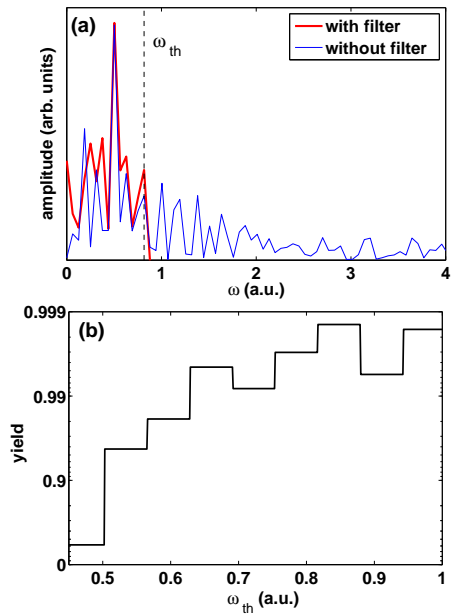}
\caption{(Color online) (a) Spectrum of the optimized field for the process $\left.|L\right>\rightarrow\left.|R\right>$
without (blue thin line) and with a spectral constraint (red thick line) at $\omega_{\rm th}=0.817$. 
(b) Occupation of target state as the function of the frequency threshold $\omega_{\rm th}$ used as the filter.
The step-like form of the curve is a consequence of the discrete Fourier transform.}
\label{fig4}
\end{figure}

After the preparation of the localized state we optimize
a transition from $\left.|L\right>$ to $\left.|R\right>$, i.e.,
a charge transfer between the quantum wells. The result of the optimization 
is summarized in Fig.~\ref{fig3}.
The optimal field having a fixed duration of $T=100$ [thin blue line in Fig.~\ref{fig4}(a)] leads to 
an extremely high yield of $0.9992$. 
In Fig.~\ref{fig3}(b) and (c) we plot the occupations of the five lowest states during
the charge-transfer process. Each of these states 
reach a maximum occupancy of more than $10\,\%$ 
during the process. The tenth lowest state still
obtains $\sim 1\,\%$ of the occupation. 
Thus, with the present length of the field,
the inclusion of several states seems to be crucial for 
the success of the optimization. Consequently, an alternative 
OCT procedure for a few-level model system (higher levels omitted) would lead
to a completely different solution field, which most likely would 
perform poorly when applied to the ``full'' system (as here) 
due to the leaking of the occupancy to higher states.~\cite{ringpaper}

Similarly to the preparation field in Fig.~\ref{fig2}(a), 
the optimized charge-transfer field in Fig.~\ref{fig4}(a) shows
abrupt peaks corresponding to high frequencies.
Hence, the field would be practically impossible to construct, e.g., with the present
pulse-shaping techniques. To relieve these limitations, we apply a spectral
constraint cutting off the high-frequency components beyond a selected threshold
frequency $\omega_{\rm th}$. The thick red line in Fig.~\ref{fig3}(a) shows the field
obtained using $\omega_{\rm th}=0.817$ ($\sim 14$ THz) in the optimization.
The Fourier spectra of both fields are shown in Fig.~\ref{fig4}(a).
Both fields have a peak at at $\omega=0.5$, which in fact corresponds to the
{\em oscillator frequency} $\omega_0$ in Eq.~(\ref{vc}).

It is interesting to note that despite the relatively strong frequency constraint 
at $\omega_{\rm th}=0.817$, leading to
a considerable smaller search space for the optimization,
the obtained yield is reduced only down to $0.9986$. This is a significant result in view 
of the fact that the original field has a large fraction of high frequencies as shown 
in Fig.~\ref{fig4}(a). Nevertheless, using a frequency filter does not considerably reduce the 
importance of higher states in the optimization: in this particular case the fifth lowest state
still gains a maximum occupancy of $\sim 10\,\%$. In any case,
further tightening of the threshold to smaller
values leads to decrease in the overlap as demonstrated in Fig.~\ref{fig4}(b).
The dependency is nonmonotonic due to numerical variation (note the logarithmic scale)
and has a step structure resulting from the discrete Fourier transform.
Below $\omega_{\rm th}\sim 0.5$ corresponding to the oscillator frequency
the fidelity collapses from $96\,\%$ to $42\,\%$. If the fluence of the field 
is increased from $0.3$ to $1$, the critical threshold remains at $0.5$, at which
the fidelity decreases from $92\,\%$ to $74\,\%$. 

\begin{figure}
\includegraphics[width=0.70\columnwidth]{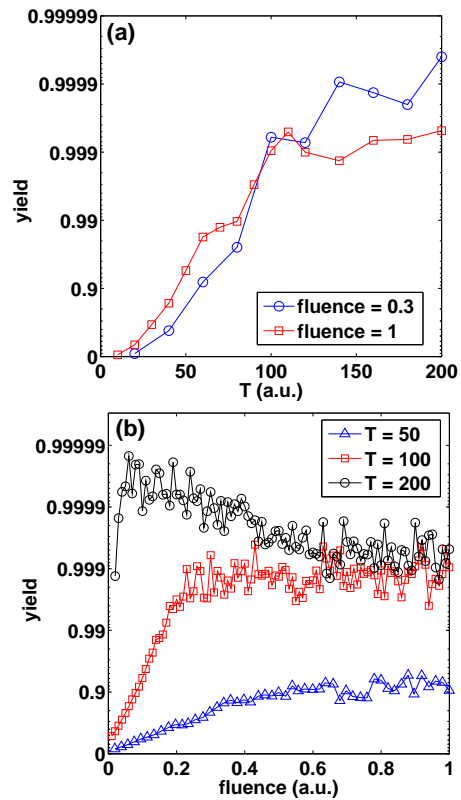}
\caption{(Color online) (a)
Obtained yield for the process $\left.|L\right>\rightarrow\left.|R\right>$
as a function of the field length.
(b) Yield for the same process as a function of the fluence
for three field lengths.
}
\label{fig5}
\end{figure}

Besides the threshold frequency, 
the main constraints in the field to be optimized
are the length and the fluence. In Fig.~\ref{fig5}(a)
we show the yield, again for the process $\left.|L\right>\rightarrow\left.|R\right>$,
as a function of the field length for fixed fluence values $E_0=0.3$ and $1$, respectively.
Both cases show some saturation around $T\gtrsim 100$ although, as expected, the smaller fluence allows
longer fields with even higher fidelities. However, increasing the yield above
$0.9999$ is difficult in this fluence range unless relatively long fields are required. 
Here, the chosen length $T=100$ seems an appropriate compromise between $T$ and the obtained yield.

Figure~\ref{fig5}(b) shows the yield as a function of the fluence for three fixed field lengths.
We remind that the fluence is a time-integrated quantity [see Eq.~(\ref{controleqs})] so that
the curves correspond to different {\em distributions} of the energy in the field. In all cases
the yield first increases exponentially with the fluence until a point of saturation is reached.
When $T=200$ the slight decrease in the overlap at fluences above $\sim 0.2$ might be due
to numerical constraints: in that regime higher and higher states (with an increasing number of nodes) are
required, and they have a finite accuracy on the numerical grid.

\begin{figure}
\includegraphics[width=0.90\columnwidth]{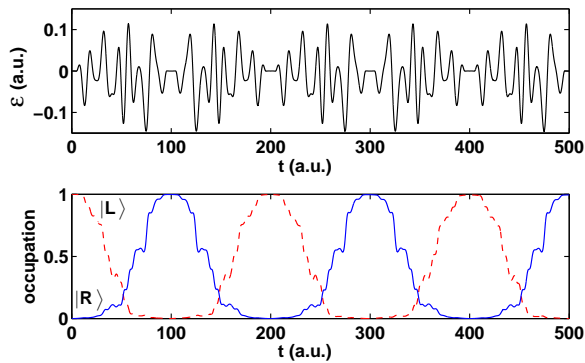}
\caption{(Color online) Optimized field (upper panel) for the five-fold charge-switch process 
$\left.|L\right>\rightarrow\left.|R\right>\rightarrow\left.|L\right>...\rightarrow \left.|R\right>$ 
(lower panel) in a double quantum dot. Here we have used a threshold frequency of $\omega_{\rm th}=0.817$ leading
to the target state occupation of $99.46\,\%$ at the end of the total five-fold process.}
\label{fig6}
\end{figure}

Finally we consider {\em sequential} charge-transfer processes by merging optimized fields together. 
For the process  $\left.|L\right>\rightarrow\left.|R\right>\rightarrow\left.|L\right>\rightarrow\ldots$ 
we combine, in turns, the optimized field $\epsilon_{L\rightarrow R}$ (see above) with its {\em time-inversion} corresponding
to $\epsilon_{R\rightarrow L}$. The combined field with a threshold frequency $\omega_{\rm th}=0.817$
is visualized in the upper panel of Fig.~\ref{fig6}.
The lower panel shows the occupations of the states $\left.|R\right>$ and $\left.|L\right>$ by solid 
and dotted lines, respectively. The final yield after the five-fold process is
$99.46\,\%$. 

A more complete view on the results of up to 100 sequential processes is given in Table~\ref{table}.
\begin{table}
\caption{Final-state occupations after $n$-fold sequential charge-switch processes obtained when merging the optimized
fields. They are compared with an estimate based on the power of the yield given by the 
original (not inverted) field (see text).}
\label{table}
  \begin{tabular}{c c c c c c c c}
  \hline
$E_0$ & $\omega_{th}$ & $n=1$ & $n=2$ & $n=5$ & $n=10$ & $n=50$ & $n=100$ \\
\hline
$0.3$ & $\infty$ & 0.9992(4) & 0.9989 & 0.9975 & 0.9961 & 0.9848 & 0.9899\\  
      &          & power law & 0.9985 & 0.9962 & 0.9924 & 0.9625 & 0.9263\\ 
  \hline
$0.3$ & $0.817$  & 0.9985(9) & 0.9977 & 0.9946 & 0.9865 & 0.9562 & 0.9589\\  
      &          & power law & 0.9972 & 0.9929 & 0.9859 & 0.9317 & 0.8681\\ 
  \hline
$0.3$ & $0.629$  & 0.9954(7) & 0.9896 & 0.9606 & 0.8874 & 0.7092  & 0.7200\\  
      &          & power law & 0.9910 & 0.9775 & 0.9556 & 0.7968 & 0.6349\\ 
  \hline
$1.0$ & $\infty$ & 0.9990(3) & 0.9984 & 0.9964 & 0.9919 & 0.9952 & 0.9863\\  
      &          & power law & 0.9981 & 0.9952 & 0.9903 & 0.9526 & 0.9074\\ 
  \hline
$1.0$ & $0.817$  & 0.9955(4) & 0.9909 & 0.9842 & 0.9837 & 0.9838 & 0.9856\\  
      &          & power law & 0.9911 & 0.9779 & 0.9563 & 0.7999 & 0.6398\\ 
  \hline
  \end{tabular}
\end{table}
We consider fluences $E_0=0.3$ and $1$ for a single process, respectively, and different threshold 
frequencies as well as the case without a filter, i.e., $\omega_{\rm th}\rightarrow\infty$.
The total yield shown in the table can be expected to (roughly) follow a power law, 
$J_{1,\rm tot}=J_{1,\rm single}^n$, where $n$ is the number of processes (charge transfers).
In this respect, the fidelity for a {\em single} transfer 
is essential for the quality of the final result.
Indeed, the computational result follows the trend of the power
law, but we find also significant differences: most importantly, 
in all cases the computational result
is better than the prediction of the power law. The most dramatic discrepancy can be
found in the last example with $E_0=1$ and $\omega_{\rm th}=0.817$, where after 100 pulses
the yield is still almost $99\,\%$, whereas the power law predicts is only $64\,\%$.
The reason behind the robustness of the yield in a sequential process is in the identity 
of the frequency components between the original and inverted fields, so that the population
``lost'' in higher states is partially attained back in the inverse process. There is, however,
no clear trend in Table~\ref{table} indicating which field parameters are particularly
favorable for robust sequential processes. Construction of such population-preserving, yet
well optimized sequential fields is a subject of future work.

We point out that a critical aspect in the feasibility of the present
approach is the sensitivity to decoherence. Typical decoherence mechanisms in 
semiconductor quantum dots are the hyperfine effects and interactions with 
optical and acoustic phonons. Their interplay and significance are largely
dependent on the external conditions in a particular device. Detailed assessment 
of these mechanisms is beyond the scope of this work. We only mention that 
typical decoherence times in semiconductor quantum dots have been measured to
be relatively large, even up to the millisecond scale,~\cite{decoherence} 
which in fact has been one of the main motivations of utilizing quantum dots in
solid-state quantum computing.~\cite{loss} In view of the time scales 
considered here (up to hundreds of picoseconds) we believe that our approach
is robust against the essential sources of decoherence, although further
analysis is in order.

\section{Summary}

Here we have numerically constructed optimal fields for charge-transfer processes
in single-electron double quantum dots. The only approximation has been the model
potential for the device, so that no truncation of eigenstates in terms of $N$-level
approximations have been used. We have found that optimal control theory provides
an efficient way to operate on desired superpositions of the eigenstates regarding
both the preparation of the localized state as well as coherent charge transfer between
the quantum wells. We have analyzed the interplay between different field constraints
including the frequency filter, fluence, and the field length. Relatively strict
frequency filters can be used without losing
the extremely high yields obtained in the processes. Combination of the optimized pulses 
can be used in sequential charge transfers while preserving the high fidelity.

\begin{acknowledgments}
This work has been funded by the Academy of Finland. 
A. P. acknowledges support by the Finnish Academy of Science and Letters, 
Vilho, Yrj\"o and Kalle V\"ais\"al\"a Foundation.
\end{acknowledgments}

\end{document}